\documentclass[twocolumn,showpacs,preprintnumbers,amsmath,amssymb]{revtex4}

\usepackage{graphicx}
\usepackage{dcolumn}
\usepackage{bm}

\begin{document}

\preprint{Phys.Rev.B}

\title{Magnetic field induced transition in a wide parabolic
well superimposed with superlattice}
\author{G. M. Gusev,$^1$ Yu.A.Pusep,$^2$ A.K.Bakarov,$^3$  A.I.Toropov,$^3$
and J. C. Portal$^{4,5,6}$} \affiliation{$^1$Instituto de
F\'{\i}sica da Universidade de S\~ao Paulo, 135960-170, S\~ao Paulo,
SP, Brazil} \affiliation{$^2$Instituto de F\'{\i}sica de S\~ao
Carlos da Universidade de S\~ao Paulo, CP 66318 CEP 05315-970, S\~ao
Carlos, SP, Brazil} \affiliation{$^3$Institute of Semiconductor
Physics, Novosibirsk 630090, Russia} \affiliation{$^4$LNCMI-CNRS,
UPR 3228, BP 166, 38042 Grenoble Cedex 9, France}
\affiliation{$^5$INSA Toulouse, 31077 Toulouse Cedex 4, France}
\affiliation{$^6$Institut Universitaire de France, 75005 Paris,
France}
\date{\today}
\begin{abstract}
We study a $Al_{x}Ga_{x-1}As$ parabolic quantum wells (PQW) with
$GaAs/Al_{x}Ga_{x-1}As$ square superlattice. The magnetotransport in
PQW with intentionally disordered short-period superlattice reveals
a surprising transition from electrons distribution over whole
parabolic well to independent-layer states with unequal density. The
transition occurs in the perpendicular magnetic field at Landau
filling factor $\nu\approx3$ and is signaled by the appearance of
the strong and developing fractional quantum Hall (FQH) states and
by the enhanced slope of the Hall resistance. We attribute the
transition to the possible electron localization in the x-y plane
inside the lateral wells, and formation of the FQH states in the
central well of the superlattice, driven by electron-electron
interaction.


\end{abstract}

\maketitle
\section{Introduction}
The multicomponent quantum Hall system, which consist of multiple
quantum wells separated by the tunneling barriers, have exhibited a
many of interesting phenomena in the strong perpendicular magnetic
field due to the interlayer electronic correlations \cite{sarma}.
The previous theoretical works suggested several possible ground
states in multilayer systems. The first class of the candidate
states is the spontaneous coherent miniband state in superlattice
(SL) quantum Hall system \cite{hanna}. This state is an analog of
interlayer coherent state at Landau filling factor $\nu=1$ in double
well structures \cite{eisenstein}. The second class of the candidate
states is the solid state phase (Wigner crystals) \cite{qiu} with
the different configurations depending on the interlayer separation.
A third type of the candidate state is staggered liquid state, which
consists of independent-layer states with unequal density
\cite{hanna}. Recently, it has been argued that another ground
state, so called dimer state, is favored for large number of layers
and small interlayer distance. The superlattice separates into pairs
of adjacent interlayer coherent states, while such coherence is
absent between layers of different pairs \cite{shevchenko}.

More recently novel three-dimensional (3D) fractional quantum Hall
states in multilayer systems have been theoretically predicted
\cite{levin}. A 3D multilayer fractional quantum Hall state with
average filling $\nu = 1/3$ per layer that is qualitatively distinct
from a stacking of weakly coupled Laughlin states was constructed
using the parton states. This new state supports gapped fermionic
quasiparticles (with charge e/3) that might propagate both within
and between the layers, in contrast to the quasiparticles in a
multilayer Laughlin state which are confined within each layer.

Despite the considerable theoretical efforts and predictions of the
many exotic broken symmetry states in quantum Hall superlattice,
such states have not yet been observed. The experimental challenge
is the fabrication of the low-disorder superlattice. However, in
refs. \cite{jo, baskey} the breakthrough idea to use the parabolic
quantum well (PQW) with a periodic modulation to produce a high
quality SL has been realized. Previously the wide PQW without SL
have been used to obtain the system with flat potential and constant
electronic density slab \cite{shayegan, sundaram}. In selectively
doped wide PQW the electrons are spatially separated from dopant
atoms and this enhances mobility and provides an opportunity to
study the  clean quasi-three dimensional system. The electron in
well screens the parabolic potential, and flat potential profile of
$\sim100\div400 nm$ width is expected. By adding the periodic
potential to this system, the clean SL can be obtained. The Fig.1 a
shows schematically empty and partially full parabolic well with
periodical and disordered SL. Indeed in Ref. 7,8 it has been
demonstrated that the mobility of electrons in PQW superimposed with
periodical superlattice is drastically enhanced in comparison with
conventional $GaAs/Al_{x}Ga_{x-1}As$ superlattices \cite{stormer,
pusep}. However, neither the coherent states in PQW with
superlattice nor other manybody effects have not yet been observed.
\begin{table*}[h!]
\label{tableI} \caption {The sample parameters. $W$ is the well
width, $n^{+}$ is characteristic bulk density of PQW, $n_s$ the
electron density, $W_{eff}$ is the effective well width, $\mu$ the
zero field mobility. $t_{z}$ is the interlayer coupling energy, and
$\delta$ is the interlayer disordered parameter.}
\begin{center}
\begin{ruledtabular}
\begin{tabular}{cccccccc}
Sample  \,\,\,  &W \,\,\,&$n^{+}$\,\,\, &$n_{s}$ &$W_{eff}$ &$\mu$
&$t_{z}$ &$\delta$\\
&($\AA$) &$(10^{16}\, cm^{-3})$ & $(10^{11}\, cm^{-2})$&$(\AA)$
&$(cm^{2}/Vs)$&(meV) &$\,\,\,$
\\  \colrule
\hline $ A $ &2400 &2.5 &3.7& 1480 &136000 & no superlattice& no superlattice\\
$B$&2400 &2.5 &3.7 &1480&200000& 1.5 & 0 \\
$C$&2400 &2.5 &3.7 &1480&186000& 1.5 & 4.73 \\
$D$&2400 &2.5 &3.7 &1480&156000& 1.5 & 8.5 \\
$E$&2400 &2.5 &3.7 &1480&209750& 1.5 & 13.5 \\
 \hline
\end{tabular}
\end{ruledtabular}
\end{center}
\end{table*}

In the present paper we report the magnetotransport measurements in
PQW with periodical and intentionally disordered short-period SL. In
strong magnetic field we observed numerous well developed plateaus
in the Hall resistance at fractional filling factors $\nu<1$ with
deep minima in the longitudinal resistance. Surprisingly, the slope
of the Hall resistance is dramatically enhanced above the critical
magnetic field, which corresponds to $\nu\approx3$. We interpret
both these effects as signaling a transition from thick slab
electronic charge distribution at $\nu>3$ to individual well
distribution at $\nu<3$ due to electron-electron interaction. We
believe that the new state has a modulation of the charge in z
direction, perpendicular to the superlattice (staggered-like state),
and at the same time electrons in lateral wells are localized by
impurities and does not contribute to the Hall and longitudinal
conductivities. The electrons in the central well of the
superlattice form FQH states which are clearly seen in experiments.
The transition is absent in PQW without superlattice.

\begin{figure}[ht!]
\includegraphics[width=9cm,clip=]{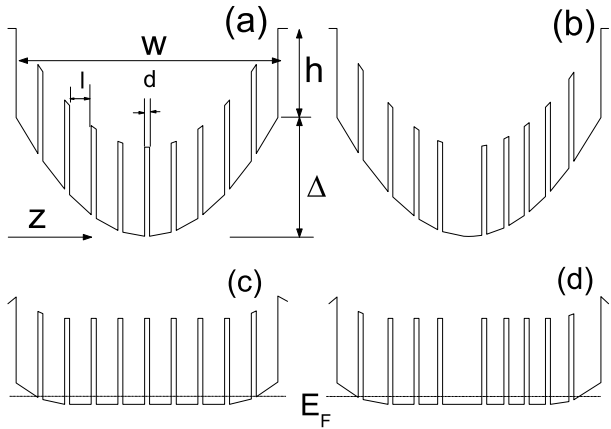}
\caption{\label{fig.1}(Color online) The profile of the conduction
band edge of an empty (a,b) and partially fill (c,d) parabolic well
with periodic (a,c) and intentionally disordered (b,d)
superlattices. $E_{F}$ is the Fermi level. }
\end{figure}

\section{Experimental results and discussions}

The samples were made from parabolic quantum well grown by molecular
-beam epitaxy. It included a $2400\AA$ -wide parabolic
$Al_{c}Ga_{c-1}As$ well with $c$ varying between $0$ and $0.29$,
bounded by undoped  $Al_{b} Ga_{b-1}As$ spacer layers with
$\delta$-Si doping on two sides \cite{gusev}. Assuming as $[001]$
the growth direction, and taking as $z=0$ the position of the pure
GaAs material, an effective harmonic potential is given by
$U=m^{\ast}\Omega^2z^{2}/2$ with $\Omega = a(2/m^{\ast})^{1/2}$ and
effective mass $m^{\ast}$, when a composition profile $c(z)=az^{2}$
is achieved. We fabricated samples with parabolic well of width
W=2400 $\AA$ and hight $h$=210 meV. The characteristic bulk density
is given by equation
$n_{+}=\frac{\Omega_{0}^{2}m^{\ast}\varepsilon}{4\pi e^{2}}$. The
effective thickness of the electronic slab can be obtained from
equation $W_{eff}=n_{s}/n_{+}$. For partially filled quantum well
$W_{eff}$ is smaller than the geometrical width of the well $W$.

In addition we produced several PQW with periodic and aperiodic
square modulation on it. The total number of 10 weakly coupled wells
was embedded in the parabolic quantum well. They consisted of wells
with thickness l = m monolayers (ML) and barriers with fixed
thickness d = 15 ML. The interlayer coupling energy $t_{z}$ = 1.5
meV was calculated according to the effective mass approximation in
the periodic superlattice with l = 65 ML.  The randomization was
achieved by a random variation of the layer thickness $m$ around the
nominal value $m$ = 65 ML according to a probability distribution
which is obtained from a Gaussian probability density for the
electron energy in the isolated well. The Gaussian is centered at
the value of the electron energy corresponding to the 65-ML well and
it is characterized by its full width at half-maximum $\Delta$
(disorder energy). The strength of the interlayer disorder is
characterized by the parameter $\delta$ = $\Delta/ t_{z}$. No
coherent interlayer tunneling is expected at $\delta > 1$.
\begin{figure}[ht!]
\includegraphics[width=9cm,clip=]{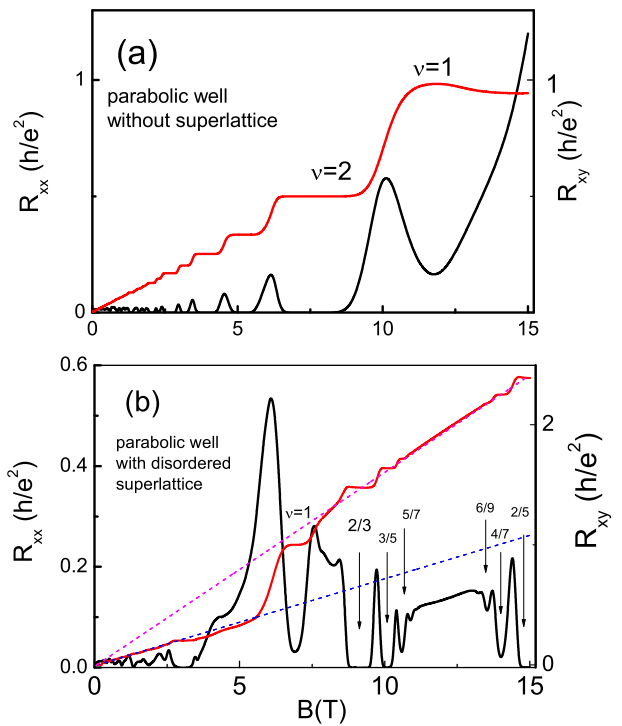}
\caption{\label{fig.2}(Color online) Longitudinal (black) and Hall
(red) resistances as functions of the perpendicular magnetic field
for a parabolic quantum well without superlattice (a)(sample A) and
with disordered superlattice (b) (sample E). Filling factors
determined from the Hall resistance are labeled.  FQH states are
marked with arrows.The dashed lines correspond to the linear
extrapolation of the low-field (blue) and high-field (magenta) Hall
resistances.}
\end{figure}
The mobility of the electron gas in our samples was
$(140-200)\times10^{3} cm^{2}$/Vs  and density - $3.7\times10^{11}
cm^{2}$.  The parameters of the samples are shown in the Table.I.
Since the effective width of the wide parabolic well is smaller then
the geometrical width, our superlattice structures have only 6-7
quantum well filled by electrons.  The test samples were Hall bars
with the distance between the voltage probes L=500 $\mu$m and the
width of the bar d=200 $\mu$m. Four -terminal resistance $R_{xx}$
and Hall $R_{xy}$ measurements were made down to 50 mK in a magnetic
field up to 15 T. The sample was immersed in a mixing chamber of a
top-loading dilution refrigerator. The measurements were performed
with an ac current not exceeding $10^{-7}-10^{-8}$ A.
\begin{figure}[ht!]
\includegraphics[width=9cm,clip=]{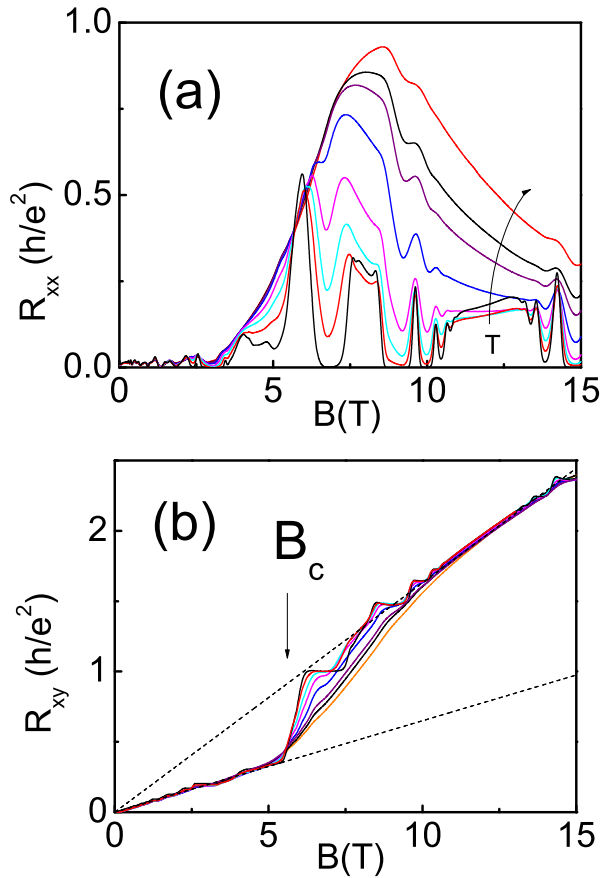}
\caption{\label{fig.3}(Color online) Longitudinal and Hall
resistances of a parabolic quantum well with disordered superlattice
 (sample E) for different temperatures T: 950 mK (red), 750 mK (black), 700 mK
(purple), 520 mK ( blue), 475 mK (magenta), 320 mK(cyan), 240
mK(red), 90 mK (black). Filling factors determined from the Hall
resistance are labeled. The dashed line corresponds to the linear
extrapolation of the low-field Hall resistance. Arrow shows the
critical magnetic field when transition occurs. }
\end{figure}

    Fig.2 shows the longitudinal and Hall resistances for conventional
PQW (a) and PQW with SL (b) measured at T=50 mK. We find the
striking difference in the properties of these two structures.
First, the conventional PQW exhibits the integer quantum Hall effect
in the strong magnetic field, as is expected for
quasi-two-dimensional system with moderate mobility \cite{gusev2},
while the PQW with disordered SL reveals well developed FQH effect
unexpected for such sample quality. From the comparison of the
strengthening of the minima at fractional filling factors with
available data for heterostructures with the same density
\cite{sajoto}, we estimate the mobility $\mu\simeq1.5-2\times10^{6}
cm^{2}/Vs$, which is almost in one magnitude higher than zero field
mobility in our samples (see Table I.).  Second, the slope of the
Hall resistance in superlattice structure is enhanced in 2 times
above the critical magnetic field $B_{c}$, which is below
corresponding total Landau filling factor $\nu=2$ (according to the
low-field Hall data) or $\nu=1$ (considering high-field Hall
resistance). Note that the electron density, and low field Hall
resistance in both samples are the same. Finally, the behaviour of
the Shubnikov-de Haas oscillations (SdH) at low magnetic field is
also changed dramatically in SL structure in comparison with
conventional PQW. We found resemblance between low field SdH and
quantum Hall  features and that from low mobility SL \cite{stormer}.
For example, the Hall resistance exhibits the quantum Hall effect at
integer filling factors per layer, as expected for multilayer Hall
system.  However, in the present paper we focus on the high field
behaviour.

The data presented in Fig.2 provide the strong evidence for the
existence of the phase transition near the total filling factor
$\nu=3$. The transition occurs in narrow interval of magnetic field
$\Delta B\sim1.1T$. Striking mobility enhancement, excess Hall
resistance, and consequently, decrease of the total electron density
can not be explained by trivial carriers freeze-out model, which is
expected to vary gradually with magnetic field. In addition, the PQW
without SL ( Fig.2a) shows the conventional quantum Hall effect
behaviour, therefore the possibility, that enhanced Hall coefficient
in superlattice structure is described by magnetic freeze out, seems
very unlikely.

  Fig.3 shows the temperature dependence of $R_{xx}$ and $R_{xy}$
in PQW superimposed with disordered SL. We may see crossover between
two distinct regimes in the magnetoresistance evolution with T. At
high temperature (T=1 K) magnetoresistance reveals the strong peak
at B=8 T, which starts to decrease with decreasing temperature, and
below $T\approx300$ mK we observe the emergence of the FQH states.
Notice that the Hall resistance demonstrates the broad transition
between different slopes. In low temperature regime $T<200$ mK the
emergent FQH states are improved, and the plateaus in the Hall
resistance, accompanied by vanishing longitudinal resistances
$R_{xx}$, become fully developed. Such crossover may indicate that
there exists competition between two different many-body states in
our system with temperature evolution. At high temperature electrons
in extreme wells are only partially localized (as we can see from
the Hall effect) and therefore, contribute to the conductivity.
Note, however, that these high temperature states cannot be simply
described by the conductivity of the parallel channels from
different independent well, since it should result in the low total
resistivity due to a dominant contribution from the high mobility
central well. We may speculate here that this state is a many-body
collective state, similar to the charge density wave state. At lower
temperature this state is destroyed because of the localization of
the electrons in the extreme quantum wells in SL, whereas the
electrons in the central well form FQH states. Below we discuss in
more details all possible ground states in multilayer quantum Hall
system.

    Finally, we should note, that magnetotransport properties of PQW
superimposed with superlattice strongly depends on the interlayer
disordered parameter $\delta$. Figure 4 shows the of $R_{xx}$ and
$R_{xy}$ traces for PQW with periodic and disordered superlattice.
We see clear similarity in the anomalous features between the traces
in Fig.2b and that from another PQW with larger parameter $\delta$
shown in Fig.2b. On the other hand, no similarity between
magnetoresistance curves has long been seen in PQW with periodic and
disordered SL. The longitudinal resistance in PQW with periodic
superlattice is dramatically increases above 12 T, and we were not
able to measure it. The insert shows the dependence of the slope of
the Hall resistance $R_{xy}/B$ on the parameter $\delta$. We see
that the the electron density in the strong magnetic field has a
tendency to freeze out with decrease of the disordered parameter.
Although not definitive, this provides some evidence that electrons
in PQW with periodic SL and in strong magnetic field are localized
by impurities in the X-Y plane and don't contribute to the Hall
conductivity.

\begin{figure}[ht!]
\includegraphics[width=9cm,clip=]{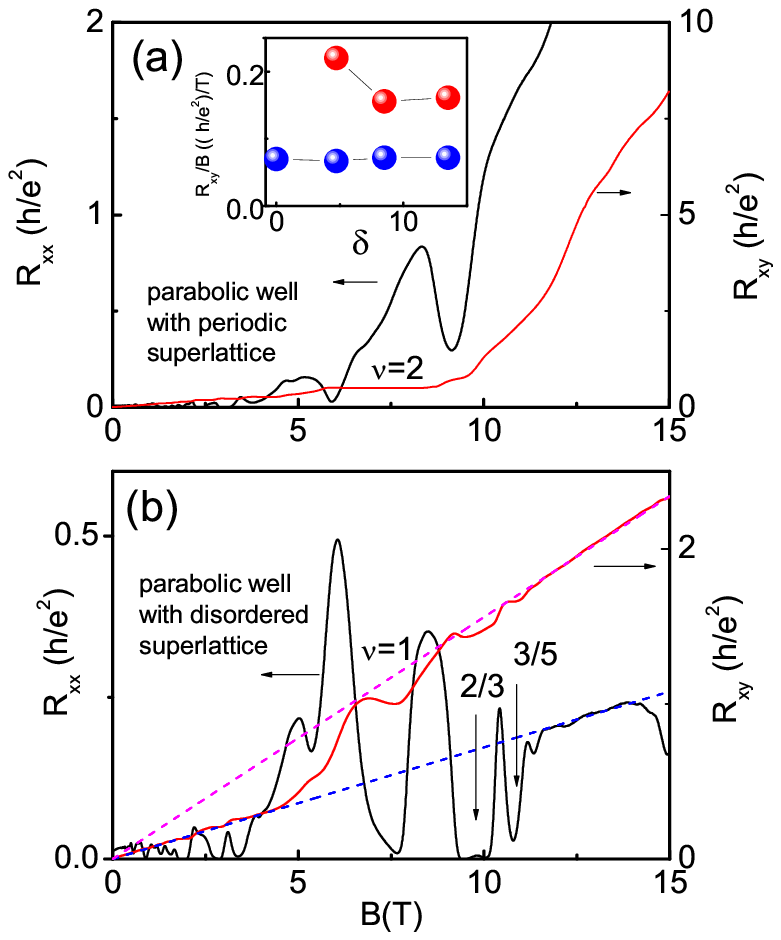}
\caption{\label{fig.4}(Color online) Longitudinal (black) and Hall
(red) resistances as functions of the perpendicular magnetic field
for a parabolic quantum well with periodic superlattice (a)( sample
B) and with disordered superlattice (b)(sample D). Filling factors
determined from the Hall resistance are labeled. The dashed lines
corresponds to the linear extrapolation of the low-field Hall (blue)
and high field (magenta) resistances. FQH states in fig.b are marked
with arrows. Insert - slope of the Hall resistance at low (blue
points) and high (red points) magnetic fields as a function of
superlattice disorder parameter.}
\end{figure}

In the following we discuss the possible origin of the transition
from multilayer quantum Hall system to the state with partially
localized electrons. Several theoretical models conclude that the
quantum Hall superlattice undergoes a phase transition
\cite{hanna,qiu,shevchenko,brey, macdonald}. The final nature of the
new manybody state depends on the parameters of the systems, such as
layer separation and interlayer coupling. Enhance of the Hall
resistance slope and emergence of the FQH states shown in Figs.2-4
make it tempting to postulate that the origin of the new ground
state is the electron-electron interaction which favors staggered
liquid state, which consists of independent-layer states with
unequal density \cite{hanna}. From this scenario it follows that the
ground state has a modulation of the charge in $z$ direction
(staggered states), which is equivalent to the formation of the
charge density wave state in three-dimensional electron system
(3DES) in the strong magnetic field \cite{kaplan, macdonald2}. At
the same time electrons are localized by impurities in $x-y$ plane.
This state is responsible for divergence of $R_{xx}$ and $R_{xy}$ in
strong magnetic field in PQW with periodic SL ( Fig.4a). In PQW with
intentionally disordered SL we break the symmetry between quantum
wells, since we introduces aperiodic square modulation in z
direction. For example, as we can see in Figure 1, the central well
is wider than the lateral wells. In this case it is very likely,
that the electronic density in the central well is higher than in
the others, therefore corresponding filling factors is sufficient to
form FQH states. At the same time, local filling factors in the
other wells are too small for FQH states and localized by the
impurities. In addition the mobility in the central well is much
higher that in the lateral wells, since the Al composition, which is
mainly responsible for electronic scattering in parabolic well, is
absent in the center of the well. Both these factors, density and
mobility, separate the properties of the individual quantum wells
and result to the formation of FQH states in the center well, and
electrons localized in X-Y plane, in the lateral wells. Note that
the alternative possible description for states in the individual
wells may be formation of the crystalline structure in x-y plane
pinned by impurities, which is equivalent to formation of the
charged Kaplan-Glasser rods parallel to the magnetic field in 3DES
case \cite{kaplan}.

 Note that transition starts after total Landau filling factor
3, below filling factor 2, while staging transition is expected at
arbitrary $\nu$ per layer. For example, in a stage-n state electrons
would occupy every n-quantum well. Therefore n=2 case corresponds to
the completely full layer with $\nu=1$, and empty layer with
$\nu=0$) \cite{hanna}. Assuming  n=2 stage-like state, we obtain 6-7
quantum well filled by electrons at low magnetic field before
transition, in agreement with our previous estimations.

It is worth noting that in the strong magnetic field after
transition the $50\%$ of electron in superlattice are localized in
the central well. Obviously, we don't see the contribution to
$R_{xx}$ and $R_{xy}$ from the nearest neighbors to the central
well. Note, that it would expected that due to the small content of
$Al$, the mobility of electrons in nearest neighbor quantum wells is
sufficient to form FQH states, which are absent in the experimental
curves. It may confirm that the new ground state is a staggered
state, and nearest neighbors to the central well are almost empty.
In this case, it is expected that only 3-4 lateral wells are
occupied by electrons. It provides the local filling factor in each
individual well $\nu^{\ast}<1/5$ at $B>9T$, which corresponds to the
Anderson localization regimes.

We should make an important point when considering Anderson
localization in the Hall insulator regime. The theoretical models
argued that in the Hall insulator phase for non interacting
\cite{kivelson} and for interacting electrons \cite{zhang}
$\sigma_{xx}=0$, $\sigma_{xy}=0$, but $\rho_{xy}=B/ne$, where n is
the total electron density, which is consistent with observations in
Anderson localization and Wigner crystal regime \cite{sajoto2,
goldman}. Therefore, neither electron localization nor
crystallization are expected  do not change the Hall resistance in
the quantum Hall system, which disagrees with our results. This
problem deserves further theoretical study.

We also remark on the similarities between Hall resistance traces in
present work and that from wide n-type ($W>4000\AA$) and p-type PQWs
without superlattices reported in previous paper \cite{zevallos}.
Authors argued that the width of the electronic and hole slabs in
PQW shrinks in the strong magnetic field due to the Hartree and
exchange correlation terms, which results to the partial charge
transfer from the PQW to the impurity level. This scenario is
supported by previous calculations of the charge density profile in
PQW in the strong magnetic field \cite{dempsey}. However,
self-consistent calculations can explain only $\sim 2\%$ decrease of
the density with B in $2400\AA$ wide PQW without SL, and fail to
explain strong increase of the density in PQW superimposed with SL.
The role of the superlattice potential in this model also remains
unclear.

\section{Conclusion}
Parabolic wells with periodic and aperiodic superlattice exhibit
strong enhancement of the Hall resistance above the critical
magnetic field. In samples with disorder SL the clear evidence of
the magnetic field induced transition is observed, which is
characterized by 2 times decrease of the density and appearance of
the FQH states unexpected for low sample quality. We attribute this
transition to separation of the electrons into two groups, one group
of electrons in the central well with highest mobility sufficient to
form FQHE states, and another group of low density and mobility
electrons in the lateral wells, which are localized in X-Y palne
and, therefore, don`t contribute to the Hall resistivity. The
transition occurs due to electron-electron interaction which favors
the redistribution of electrons in the individual wells, very likely
with unequal density (staging transition).
\section{Acknowledgments}
 Support of this
work by FAPESP, CNPq (Brazilian agencies) is acknowledged. We thank
O. E.Raichev for illuminating discussions.

\end{document}